\documentclass[ %
reprint,
amsmath,amssymb, 
aps,
prl,
]{revtex4-2} 
\usepackage{graphicx, color}

\usepackage{dcolumn}
\usepackage{bm}
\usepackage{booktabs}
\usepackage[colorlinks,urlcolor=blue,linkcolor=blue,anchorcolor=blue,citecolor=blue]{hyperref} 

\usepackage{amsmath,amsfonts,bm}

\def\eqref#1{equation~\ref{#1}}

\def\1{\bm{1}}

\def\va{{\bm{a}}}

\def\vl{{\bm{l}}}

\def\vx{{\bm{x}}}

\def\vepsilon{{\bm{\epsilon}}}

\def\mA{{\bm{A}}}

\def\mI{{\bm{I}}}

\def\mL{{\bm{L}}}

\def\mX{{\bm{X}}}

\DeclareMathAlphabet{\mathsfit}{\encodingdefault}{\sfdefault}{m}{sl}
\SetMathAlphabet{\mathsfit}{bold}{\encodingdefault}{\sfdefault}{bx}{n}

\def\gL{{\mathcal{L}}}
\def\gM{{\mathcal{M}}}
\def\gN{{\mathcal{N}}}

\def\sR{{\mathbb{R}}}

\newcommand{\E}{\mathbb{E}}

\usepackage{graphicx,color}
\usepackage{dcolumn}
\usepackage{bm}
\usepackage{booktabs}
\usepackage{float}
\usepackage{algorithm}
\usepackage{algorithmic}

\begin{document}
\preprint{APS/123-QED}
\title{
InvDesFlow: An AI-driven materials inverse design workflow to explore possible high-temperature  superconductors}
\author{Xiao-Qi Han$^{1,2}$}
\thanks{These authors contributed equally to this work.}
\author{Zhenfeng Ouyang$^{1,2}$}
\thanks{These authors contributed equally to this work.}
\author{Peng-Jie Guo$^{1,2}$}
\author{Hao Sun$^{4}$}
\author{Ze-Feng Gao$^{1,2}$}
\email{zfgao@ruc.edu.cn}
\author{Zhong-Yi Lu$^{1,2,3}$}
\email{zlu@ruc.edu.cn}
\affiliation{1. School of Physics and Beijing Key Laboratory of Opto-electronic Functional Materials $\&$ Micro-nano Devices. Renmin University of China, Beijing 100872, China}
\affiliation{2. Key Laboratory of Quantum State Construction and Manipulation (Ministry of Education), Renmin University of China, Beijing 100872, China}
\affiliation{3. Hefei National Laboratory, Hefei 230088, China}
\affiliation{4. Gaoling School of Artificial Intelligence, Renmin University of China, Beijing, China}
 
\date{\today}
\begin{abstract}
The discovery of new superconducting materials, particularly those exhibiting high critical temperature ($T_c$), has been a vibrant area of study within the field of condensed matter physics. Conventional approaches primarily rely on physical intuition to search for potential superconductors within the existing databases. However, the known materials only scratch the surface of the extensive array of possibilities within the realm of materials.  Here, we develop InvDesFlow, an AI-driven materials inverse design workflow that integrates deep model pre-training and fine-tuning techniques, diffusion models, and physics-based approaches (e.g., first-principles electronic structure calculation) for the discovery of high-$T_c$ superconductors. Utilizing InvDesFlow, we have obtained 74 thermodynamically stable materials with critical temperatures predicted by the AI model to be $T_c \geq$ 15 K based on a very small set of samples. Notably, these materials are not contained in any existing dataset. Furthermore, we analyze trends in our dataset and individual materials including B$_4$CN$_3$ (at 5 GPa) and B$_5$CN$_2$ (at ambient pressure) whose $T_c$s are 24.08 K and 15.93 K, respectively.  We demonstrate that AI technique can discover a set of new high-$T_c$ superconductors, outline its potential for accelerating discovery of the materials with targeted properties.
\end{abstract}


\maketitle

\textit{Introduction.}
Superconducting materials have numerous applications in modern society since it was discovered~\cite{onnes1911resistance}, particularly in magnetic resonance imaging~\cite{lvovsky2013novel}, fueling advances in nuclear fusion technology~\cite{bruzzone2018high}. Superconductor-based devices are demonstrating potential for achieving scalable quantum information processors, advanced sensors, and efficient communication systems~\cite{mirhosseini2020superconducting,gambetta2017building,degen2017quantum}. Many of these devices use conventional Bardeen-Cooper-Schrieffer (BCS) superconductors~\cite{bardeen1957theory}, which demand costly helium-based cooling. Hence, searching superconductors with high superconducting critical temperature ($T_c$) is vital for propelling technological progress in these dynamic areas. 

Over the past decade, substantial advancements have been achieved in searching high-$T_c$ superconductors. For example, a superconducting transition with $T_c$ $\sim$ 36 K was experimentally observed in high-pressured Scandium, which is the highest record for elemental superconductors~\cite{PhysRevLett.130.256002}. The discovery of superconductivity in bilayer La$_3$Ni$_2$O$_7$ under pressure raises superconducting $T_c$ of nickelates to the liquid-nitrogen temperature zone~\cite{Sun-nature}. And lots of theoretical work predicted superconductivity in hydrides~\cite{duan2014pressure,doi:10.1073/pnas.1704505114,PhysRevLett.119.107001,PhysRevB.104.L100504,drozdov2019superconductivity,PhysRevB.104.054501}, where superconductivity in H$_3$S under pressure was experimentally confirmed~\cite{drozdov2015conventional}. 

Recently, a series of AI-driven inverse design methodologies for materials have emerged~\cite{AImaterials-review}, such as MatAltMag~\cite{AI-altermagnetic} which is an AI-based search engines to accelerate discovery of altermagnetic materials. Meanwhile, the generative AI models were used to identify the new conventional superconducting materials~\cite{choudhary2022designing,wines2023high,stanev2018machine,zhang2022machine,menon2022generative,roter2020predicting,simclp}. This paradigm shift demonstrates the transformative potential of artificial intelligence in accelerating the exploration of functional materials with targeted properties through systematic data-driven approaches.
Wines et al~\cite{wines2023inverse}. have employed crystal diffusion variational auto-encoder~(CDVAE)~\cite{xie2022crystal} to generate data based on the JARVIS-DFT database~\cite{choudhary2020joint}, subsequently employing the atomistic line graph neural network (ALIGNN)~\cite{alignn2021} for $T_c$ forecasting. Using high-throughput density functional theory (DFT) calculations, 34 dynamically stable 2D superconductors with $T_c$ $\ge$ 5 K from over 1000 candidates in the JARVIS-DFT database were identified~\cite{wines2023high}. Moreover, Choudhary K~\cite{choudhary2022designing} leveraged electron-phonon coupling (EPC) calculations, assistanted by deep-learning models for efficient prediction of superconducting properties, to identify 105 conventional superconductors with $T_c \ge$ 5 K from a pre-screened set of 1736 materials. 
Significant progress has also been made in using machine learning to search for superconducting hydrides~\cite{supercon50,Sanna2024,PhysRevLett.132.166001} and carbide superconductors~\cite{Geng2023}. Based on InvDesFlow, we have also identified a candidate material, Li$_2$AuH$_6$, with a superconducting critical temperature of approximately 140 K~\cite{ouyang2025strong}.
Although numerous studies have highlighted the application of machine learning in this field, these approaches primarily rely on chemical formulas or searches based on existing datasets. They often lack the intricate atomic structure details crucial for understanding superconducting behavior and are limited in exploring crystal materials beyond known databases. To truly advance the discovery of new superconductors, it is essential to incorporate detailed structural information and broaden the scope beyond existing data. So far, the conventional methods (such as elemental substitution or physical insight) have limited success in finding new high-$T_c$ superconductors among the existing data. The rise of Al technology brings a transformative approach, potentially reshaping our path to solving this challenge.

In this work, we developed InvDesFlow, an AI-driven inverse design of materials workflow to explore high-$T_c$ BCS superconductors, integrating diffusion model, formation energy prediction model, ALIGNN, pre-training and fine-tuning technique, atom docking based on pre-trained model, active learning technique, and physics-based methods (\emph{e.g.}, first-principles electronic structure calculations), and meanwhile sufficiently incorporating detailed structural information. Leveraging a limited dataset of high-$T_c$ BCS superconductors~(105 superconductors with $T_c \ge$ 5 K~\cite{choudhary2022designing}), we have obtained 74 thermodynamically stable materials exhibiting critical temperatures predicted by the AI model to be $T_c \ge$ 15 K. Furthermore, we analyze trends in our results, focusing on specific materials such as B$_4$CN$_3$ and B$_5$CN$_2$, with $T_c$ of 24.08 K and 15.93 K, respectively. InvDesFlow stands out for its unique capability to obtain crystal structures absent from the existing material databases, effectively pioneering new avenues in the quest for high-$T_c$ superconductors. Its adaptability allows itself to be tailored for a diverse array of functional materials, each with specific desired properties, thereby greatly expanding its utility across the field of materials science.

\begin{figure*}[tp!]
\centering  
		\includegraphics[width=1.0\linewidth]{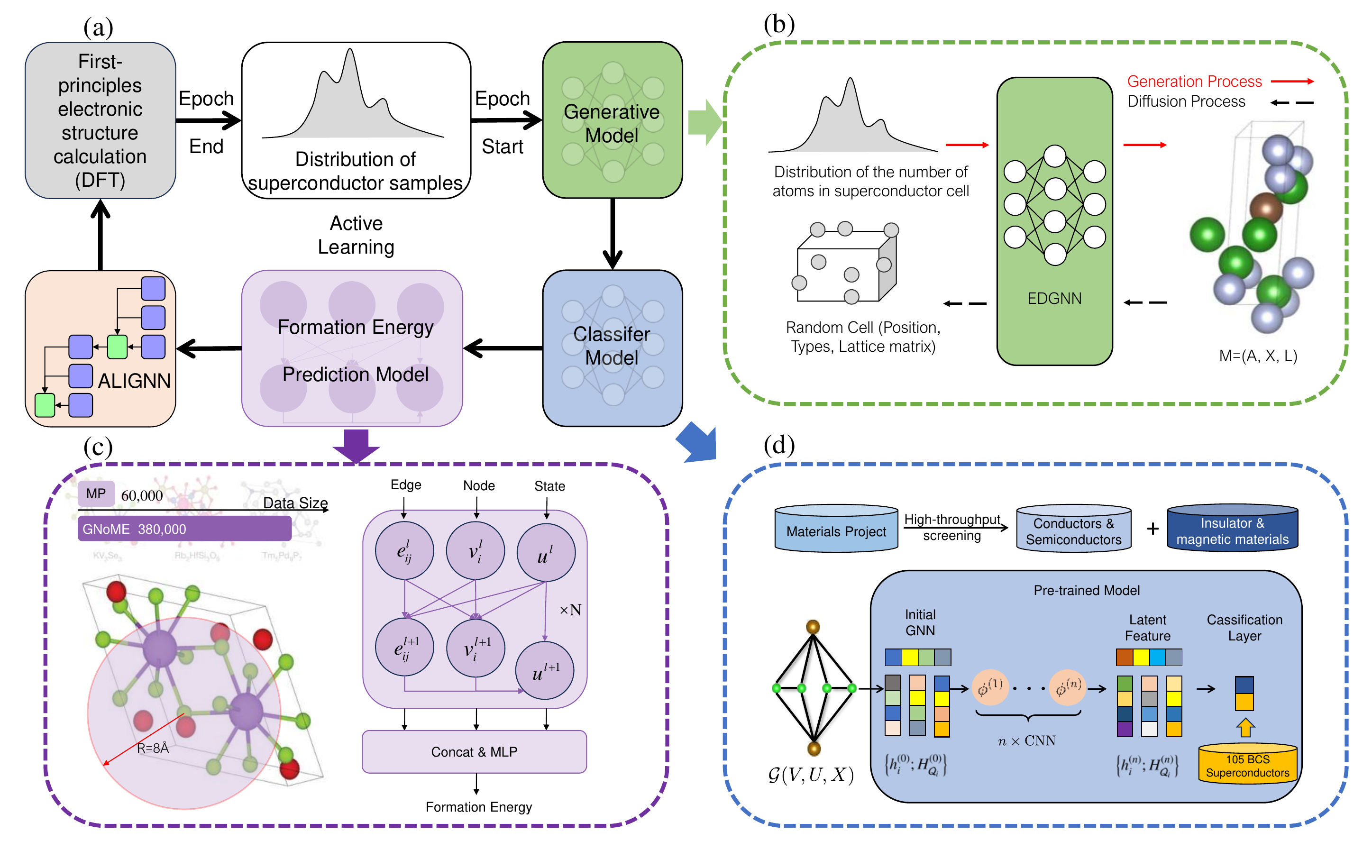}
  \caption{
  (a) The Proposed InvDesFlow framework. AI-accelerated discovery of high-$T_c$ superconductors includes generative model for predicting crystal structures, pre-trained model for superconductivity classification, formation energy prediction model, screening model for superconducting transition temperature prediction, and validation using DFT calculation. 
(b) Symmetry-constrained crystal generation model. The generation of superconducting crystals defines two Markov processes: the black arrows represent the gradual addition of noise to a BCS superconducting crystal, resulting in a random unit cell, while the red arrows indicate the gradual denoising from a prior atomic distribution to generate the original superconducting crystal structure. \textcolor{black}{The structures predicted by the generative AI have not yet converged in terms of energy and forces, requiring further post-processing. Here, we use DPA2~\cite{zhang2023dpa} to predict the interatomic potentials (at DFT accuracy) and employ the atomic simulation environment~\cite{ase-jcp} to simulate the structural relaxation.}
(c) Formation Energy Prediction Model. \textcolor{black}{A lower formation energy of a material indicates that its constituent elements adopt the lowest-energy configuration arrangement, which generally implies thermodynamic stability. In this step, we predict the formation energies of structurally optimized materials for subsequent screening of promising candidates.} This model covers the sources of training data for formation energy prediction, crystal data representation using atomic graphs with an 8 \text{Å} cutoff radius, and the interactions between node, edge, and global state representations within the model’s architecture. 
(d) Superconducting Classification Model. This model includes a high-throughput screening process for pre-training and fine-tuning data, along with a graph auto-encoder architecture based on a graph neural network for superconductivity classification.
}
  \label{fig:framwork}
\end{figure*}

\textit{Workflow overview.}

\textit{Symmetry-constrained crystal generation model.}
In crystal structures (Fig~\ref{fig:framwork} (b)), the atoms exhibit a periodic distribution, with the smallest repeating unit being the unit cell, denoted by $\gM$, which can be represented as $\gM=(\mA, \mX, \mL)$. Here, $\mA=[\va_1, \va_2, ..., \va_N]\in\sR^{h\times N}$ denotes the atomic types within the unit cell, $\mX=[\vx_1, \vx_2,...,\vx_N]\in\sR^{3\times N}$ represents the Cartesian coordinates of each atom, and $\mL=[\vl_1, \vl_2, \vl_3]\in\sR^{3\times 3}$ is the lattice matrix used to describe the periodicity of the crystal. We employed an ab initio crystal generation approach to generate superconducting crystal structures. Specifically, this involves generating a superconducting crystal $\gM$ from a given number of atoms $N$ within the unit cell, with a sampling distribution defined as:
\begin{equation}
    p(\gM,N)=p(N)p(\gM|N),
\end{equation}
where $N$ remains unchanged during the generation process. The distribution $p(N)$ is calculated from the training set, while $p(\gM|N)$ is generated based on the model. 
Standard denoising diffusion probabilistic model~(DDPM)~\cite{ho2020denoising} can be used to generate $\mL$ and $\mA$, and their loss functions take the same form as:
\begin{equation}
    \gL_{\mL / \mA}=\E_{\vepsilon\sim\gN(0,\mI)}[\|\vepsilon - \hat{\vepsilon}_{\mL / \mA}(\gM_t,t)\|_2^2].
\end{equation}
The denoising terms $\hat{\vepsilon}_\mL(\gM_t,t)$ and $\hat{\vepsilon}_\mA(\gM_t,t)$ are predicted by an equivalent denoising graph neural networks (short as EDGNN Fig~\ref{fig:framwork} (b)), and $\gN(0,\mI)$ represents a standard normal distribution, just like MatterGen~\cite{Zeni2025} and DiffCSP~\cite{jiao2024crystal}. Given the periodicity of $\mX$, it is generated using a score-matching based framework ~\cite{song2021Score-based}. For further details, \textcolor{black}{see SM Sec.1}.  In diffusion generative models, the denoising terms correspond to the noise components that the model aims to predict during the reverse diffusion process. EDGNN, as a graph neural network, is designed to estimate these denoising terms, with its performance optimized through the joint training of multiple loss functions. Utilizing 105 BCS superconductors~\cite{choudhary2022designing}, we trained the model to generate novel crystal structures, excluding those in the training set and with overlapping compositions in the Materials Project (MP) database~\cite{jain2020materials}. Since generative models often produce non-ground-state structures, we performed geometry optimization using the neural networks atomic simulation environment~\cite{ase-jcp} and L-BFGS algorithm~\cite{liu1989limited} to refine the generated structures.

\textit{Superconducting classification model.}
Initially, we extract 144,595 crystal data entries from the MP database~\cite{jain2020materials}. We first classified the materials into two groups: magnetic and non-magnetic. Subsequently, we refined the non-magnetic category into conductors, semiconductors, and insulators. Then, we designated insulators and magnetic materials as negative samples, and conductors and semiconductors as positive samples, as illustrated in Fig~\ref{fig:framwork} (d). The model is based on a pre-trained graph neural network (GNN) that utilizes material crystal structure information to predict materials~\cite{xie2018crystal, AI-altermagnetic}, consists of a graph convolutional network encoder and a decoder that reconstructs the graph features based on optimal transport theory (see Fig~\ref{fig:framwork} (d) and \textcolor{black}{SM Sec.8}). To obtain hidden layer representations related to superconductivity, we pre-trained the model using the positive samples. During the fine-tuning stage, we employed the pre-trained encoder and used up-sampling techniques to balance the number of the BCS superconductors and negative samples for binary classification model. Subsequently, we obtained the classifier model that achieved a discrimination success rate of $99.04\%$ for the 105 BCS superconductors. Utilizing this model , we evaluated the candidate structures generated by the generative model.

\textit{Formation energy prediction model.}
To further assess the stability of potential superconductors, we predict the formation energy of crystals as an indicator of their stability (Fig~\ref{fig:framwork} (c)). 
The AI algorithms like CGCNN~\cite{xie2018crystal} and SchNet~\cite{SchNet2017}, while fast, lack the required precision for formation energy predictions. Inspired by MEGNET~\cite{chen2019graph}, 
we trained the model using 380,000 crystal structures from GNoME~\cite{merchant2023scaling} and 60,000 crystal structures (from Materials Project)~\cite{jain2020materials}. Next, we increased the cutoff radius for constructing atomic graphs from 5\text {Å} to 8\text {Å}, enabling the model to capture more long-range atomic interactions to more accurately simulate atomic interactions. 
Recognizing the direct correlation between crystal formation energy and atomic bonding strength, we have incorporated eight new atomic features into our prediction model. This enhancement offers a more comprehensive representation of crystal data, as elaborated in the \textcolor{black}{SM Sec.9}.
The original MEGNET benchmark reported a mean absolute error (MAE) of 28 meV per atom, while our improved model achieved the same level of accuracy as GNoME, with an MAE of 21 meV per atom, despite GNoME not providing details of the algorithm.
Since we are particularly interested in high-$T_c$ superconducting materials, we used ALIGNN~\cite{alignn2021} to predict the superconducting transition temperatures of these materials and applied a 15 K threshold, resulting in 74 candidate high-$T_c$ superconductors. After performing computational complexity estimations for these 74 candidate materials, we selected three representative materials~(B$_5$CN$_2$ ,B$_4$CN$_3$ and B$_4$C$_3$N) for detailed DFT calculations and obtained two stable superconductors.

\begin{figure}[tp!]
  \includegraphics[width=8.6cm]{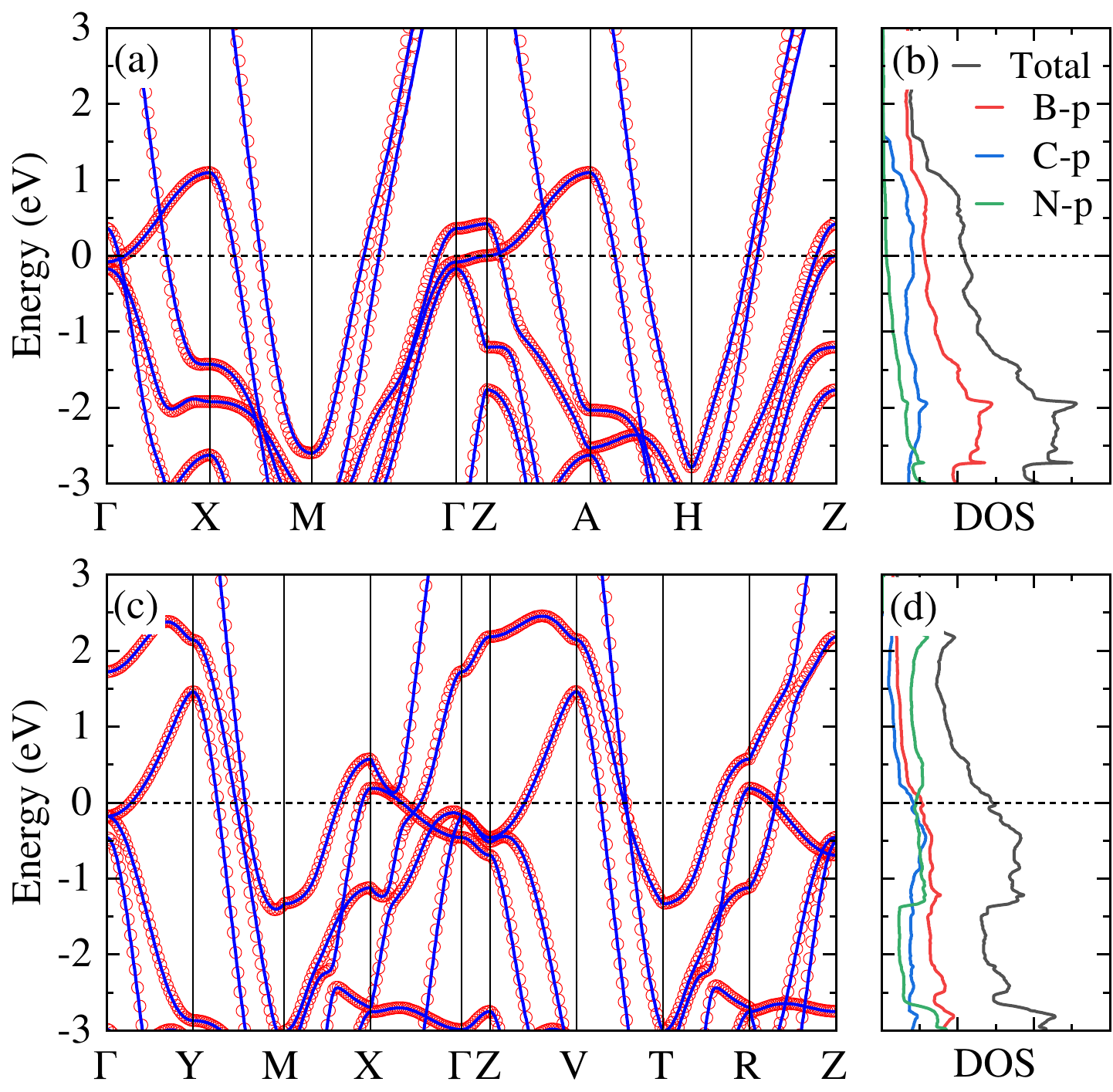}
  \caption{(a)-(b) Electronic structure and DOS of B$_5$CN$_2$ at ambient pressure. (c)-(d) Electronic structure and DOS of B$_4$CN$_3$ under 5 GPa. The blue soild lines and red circles represent the bands obtained by DFT and Wannier projection, respectively. The Fermi level is set to be zero.}
  \label{fig:band}
\end{figure}

\textit{Predicted high-$T_c$ materials.}
By performing the DFT calculations, we studied the electronic structure, phonon properties, and EPC of B$_5$CN$_2$ and B$_4$CN$_3$ (See \textcolor{black}{SM Sec.1-5} for crystal structures and additional results). In Fig.~\ref{fig:band}, we show the band structure of B$_5$CN$_2$ and B$_4$CN$_3$~(5 GPa). The results of DFT calculations and Wannier projection show good consistence and suggest that B$_5$CN$_2$ and B$_4$CN$_3$~(5 GPa) are metallic. The atomic-orbital resolved density of states~(DOS) shows that the 2$p$ orbitals of B, C, and N atoms mainly contribute the Fermi surfaces. 

\begin{figure}[thp]
  \includegraphics[width=8.6cm]{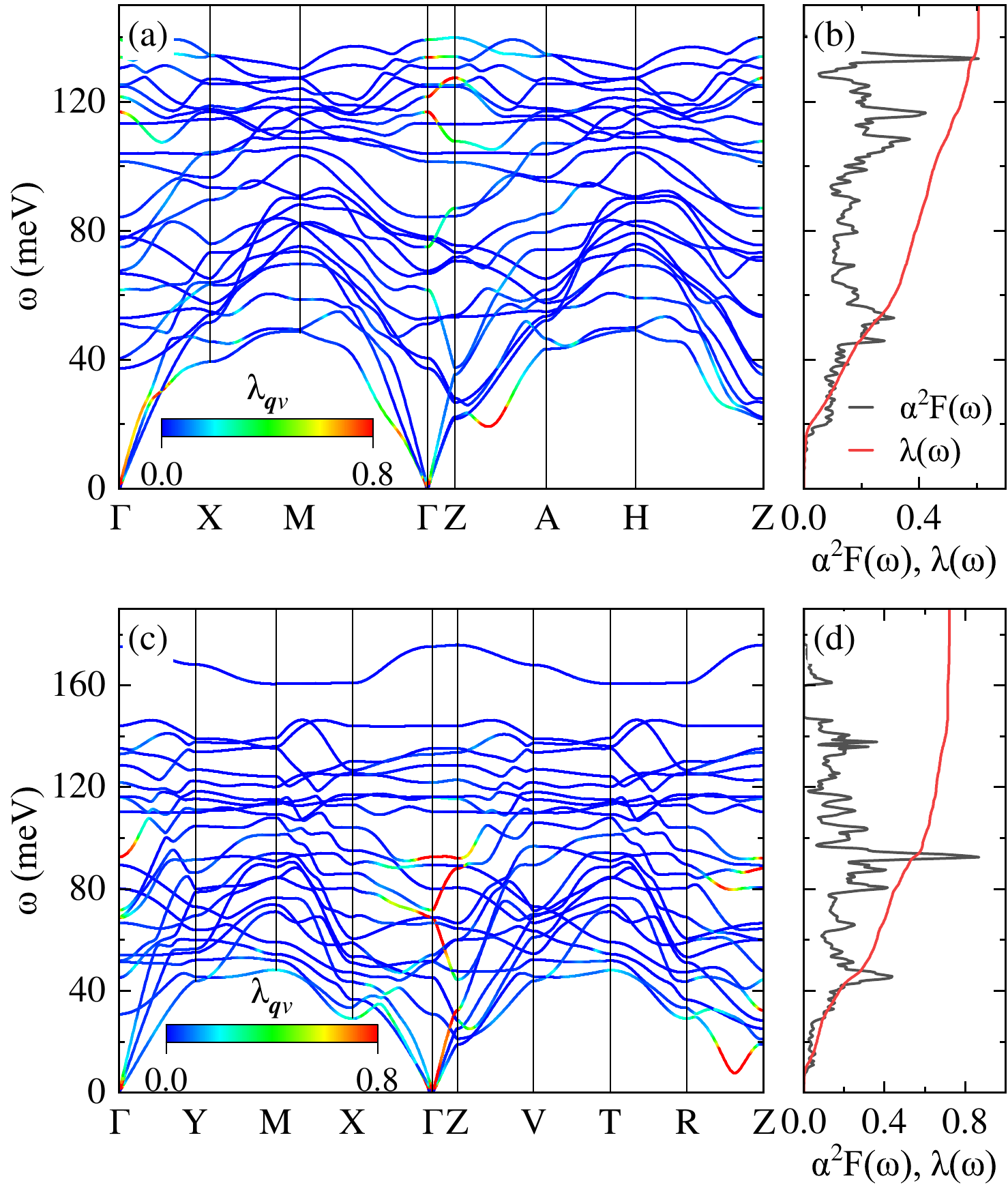}
  \caption{(a)-(b) Phonon spectrum with a color representation of $\lambda_{q\nu}$, Eliashberg spectral function $\alpha^2F(\omega)$, and accumulated EPC constant $\lambda(\omega)$ for B$_5$CN$_2$ at ambient pressure. (c)-(d) Phonon spectrum with a color representation of $\lambda_{q\nu}$, Eliashberg spectral function $\alpha^2F(\omega)$, and accumulated EPC constant $\lambda(\omega)$ for B$_4$CN$_3$ under 5 GPa. The scale of $\alpha^2F(\omega)$ is omitted.}
  \label{fig:phonon}
\end{figure}

Next, we investigate the dynamical stability of B$_5$CN$_2$ and B$_4$CN$_3$. At ambient pressure, we find that B$_5$CN$_2$ is dynamically stable, while B$_4$CN$_3$ shows a maximum imaginary-frequency phonon of $\sim$ $-7.7$ meV along the $R$-$Z$ path. By applying pressure of 5 GPa, the imaginary phonon of B$_4$CN$_3$ disappears. Hence, we show the phonon spectrum of B$_5$CN$_2$ and B$_4$CN$_3$ (5 GPa) in Figs.~\ref{fig:phonon}(a) and (c) and further study the EPC of these two materials. The calculated Eliashberg spectral function $\alpha^2F(\omega)$ and accumulated EPC constant $\lambda(\omega)$ are exhibited in Figs.~\ref{fig:phonon}(b) and (d). And the mode-resolved $\lambda_{q\nu}$ is added in the phonon spectrum. The EPC constants $\lambda$ of B$_5$CN$_2$ and B$_4$CN$_3$ (5 GPa) are integrated to be 0.61 and 0.72, respectively. Using the McMillan-Allen-Dynes formula~\cite{PhysRevB.6.2577,PhysRevB.12.905}
\begin{equation}
    T_c=\frac{\omega_{log}}{1.2}\exp[\frac{-1.04(1+\lambda)}{\lambda(1-0.62\mu^*)-\mu^*}],
\end{equation}
where $\omega_{\text{log}}$ is the logarithmic average frequency. For a detailed definition, please refer to the \textcolor{black}{SM Sec.1}. The superconducting $T_c$ of B$_5$CN$_2$ and B$_4$CN$_3$ (5 GPa) are estimated to be 15.93 K and 24.08 K, respectively, when the Coulomb pseudopotential $\mu^*$ is set to 0.1.

\textit{Discussion.}
Recently, several studies have utilized generative models to explore high-$T_c$ superconductors~\cite{choudhary2022designing,wines2023inverse, wines2023high, zhang2022machine}. Wines D~\cite{wines2023inverse} employed CDVAE to generate data on the JARVES-DFT dataset~\cite{choudhary2020joint}, subsequently employing the ALIGNN~\cite{alignn2021} for $T_c$ forecasting. Compared to existing methods, InvDesFlow has seen improvements in three aspects. Firstly, our method is capable of effective generation based on a few positive samples~(\emph{i.e.}, 105 samples with $T_c\ge$5 K). Unlike CDVAE~\cite{xie2022crystal}, which randomly generates chemical formulas before predicting structures, our approach directly generates structures. By directly generating structural configurations, our method adeptly navigates the spatial intricacies of superconductors, facilitating the genesis of plausible chemical entities. Secondly, we have integrated a sophisticated post-processing phase employing the DPA-2 model~\cite{zhang2023dpa} for atom docking. The discussion on DPA-2 achieving structure relaxation with DFT-level accuracy can be found in the \textcolor{black}{SM Sec.7}. This step meticulously circumvents atomic clashes, refines bond lengths to more rational values, and guarantees the equilibrium of forces exerted on each atomic constituent. 

\begin{table}[t]
\centering
\begin{tabular}{l|cccc}
\hline
\textbf{Model} & \textbf{FE} & \textbf{Ehull} & \textbf{EF} & \textbf{DS} \\
\hline
CDVAE~\cite{xie2022crystal} & $\checkmark$ & $\times$ & $\times$ & $\times$ \\

Con-CDVAE~\cite{concdvae} & $\checkmark$ & $\times$ & $\times$ & $\times$ \\

DiffCSP~\cite{jiao2024crystal} & $\checkmark$ & $\times$ & $\times$ & $\times$ \\

CrystaLLM~\cite{Antunes2024} & $\checkmark$ & $\times$ & $\times$ & $\times$ \\

MatterGen~\cite{Zeni2025} & $\checkmark$ & $\times$ & $\times$ & $\times$ \\
\hline
InvDesFlow & $\checkmark$ & $\times$ & $\checkmark$ & $\times$ \\
\hline
\end{tabular}

\caption{Stability evaluation of materials generated by different AI models, including thermodynamic stability (formation energy less than 0), synthesizability (hull energy equal to 0), atomic forces (atomic interactions predicted to be within the precision of DFT-optimized structures), and dynamical stability (absence of imaginary frequencies in phonon spectra). Abbreviations used: FE for formation energy, Ehull for convex hull energy, EF for atomic forces and energy, and DS for dynamical stability.}
\end{table}

The stability of structures generated by AI methods is discussed from multiple aspects, including thermodynamic stability, Ehull, energy and interatomic forces, and dynamical stability. Current leading generative AI algorithms, such as , MatterGen~\cite{Zeni2025}, DiffCSP~\cite{jiao2024crystal}, CDVAE~\cite{xie2022crystal}, Con-CDVAE~\cite{concdvae}, and CrystaLLM~\cite{Antunes2024}, generate new crystal structures by learning from the distribution of crystals in the training set. Due to specific chemical compositions typically satisfying the formation energy \textcolor{black}{$\textless$} 0 criterion, these generated materials demonstrate good thermodynamic stability. However, the Ehull of directly generated materials is usually not equal to 0, requiring further DFT calculations to confirm their stability. Additionally, current AI algorithms cannot directly achieve the precision of DFT structural relaxation in predicting interatomic forces and energies, so the generated structures often require post-processing to optimize atomic interactions. Regarding dynamical stability, typically referring to whether the material's phonon spectrum contains imaginary frequencies, this is an aspect that generative AI methods currently cannot directly address. Therefore, further optimization and validation are necessary to ensure the complete stability of the generated materials.

Existing methods often predict superconducting transition temperatures without aforehand confirming the materials' superconductivity, which is problematic. We address this by introducing a superconducting classification model. We improved the formation energy prediction model under GNoME~\cite{merchant2023scaling}, increasing its precision from 28 meV to 21 meV. Lastly, active learning progressively expands the chemical space of high-$T_c$ superconducting materials in iterative reinforcement generative learning. These refinements enhance our method's effectiveness and establish a new standard for exploring and predicting high-$T_c$ superconductors, leading to potential breakthrough in materials science and technology.

\textcolor{black}{Although InvDesFlow demonstrates potential in the discovery of high-temperature superconducting (HTS) materials, its main limitation lies in the limited availability of high-quality HTS data. The current generative model relies on 105 HTS data points, and the ALIGNN $T_c$ prediction model is also based on this dataset. As a result, the 74 newly generated candidate materials may inherit features from the original data, with most containing nitrogen atoms and $T_c$ values concentrated between 15-20 K.To overcome this limitation, we plan to expand the dataset by incorporating external HTS data and continuously optimize the $T_c$ prediction model, particularly by introducing more diverse chemical compositions and crystal structures. Additionally, through iterative active learning, we aim to explore potential HTS materials not included in the existing databases. We believe that with the expansion of the dataset and optimization of the models, InvDesFlow will be able to overcome its current limitations and further advance the discovery of high-temperature superconducting materials.}

In conclusion, InvDesFlow integrates a suite of advanced methodologies, including generative model, formation energy prediction model, pre-training and fine-tuning strategy, ALIGNN, and first-principles electronic structure calculations. This AI search engine has not only predicted 74 superconducting material candidates (see \textcolor{black}{SM Sec.6}) with $T_c>$15 K based on a modest set of positive samples~(105 samples with $T_c\ge$5 K), but also identified two ideal high-$T_c$ candidates: B$_5$CN$_2$ ($T_c$=15.93 K) and B$_4$CN$_3$ ($T_c$=24.08 K). Notably, this engine is capable of discovering crystal structures that are not yet documented in existing material dataset, thereby opening up new horizons in the search for high-$T_c$ superconductors. Moreover, the AI search engine's flexibility expands its utility in materials science for various functional materials.

\textit{Open Data and Code Availability.}
In order to support the development of the field of materials inverse design and to enable readers to replicate our work, we have made all the data and code publicly available. The code repository can be accessed at \url{https://github.com/xqh19970407/InvDesFlow}.

\textit{Acknowledgments.}
This work was financially supported by the National Natural Science Foundation of China (Grant No.62476278, No.12434009, and No.12204533), the National Key R\&D Program of China (Grants No. 2024YFA1408601), and the Innovation Program for Quantum Science and Technology (Grant No. 2021ZD0302402). Computational resources have been provided by the Physical Laboratory of High Performance Computing at Renmin University of China.

\bibliographystyle{IEEEtran}

\begin{thebibliography}{}
\providecommand{\url}[1]{#1}
\csname url@samestyle\endcsname
\providecommand{\newblock}{\relax}
\providecommand{\bibinfo}[2]{#2}
\providecommand{\BIBentrySTDinterwordspacing}{\spaceskip=0pt\relax}
\providecommand{\BIBentryALTinterwordstretchfactor}{4}
\providecommand{\BIBentryALTinterwordspacing}{\spaceskip=\fontdimen2\font plus
\BIBentryALTinterwordstretchfactor\fontdimen3\font minus \fontdimen4\font\relax}
\providecommand{\BIBforeignlanguage}[2]{{%
\expandafter\ifx\csname l@#1\endcsname\relax
\typeout{** WARNING: IEEEtran.bst: No hyphenation pattern has been}%
\typeout{** loaded for the language `#1'. Using the pattern for}%
\typeout{** the default language instead.}%
\else
\language=\csname l@#1\endcsname
\fi
#2}}
\providecommand{\BIBdecl}{\relax}
\BIBdecl

\end{thebibliography}


\begin{thebibliography}{99}\footnotesize
\itemsep=-1pt plus.2pt minus.2pt

\bibitem {onnes1911resistance} Onnes H K 1911 {\it Commun Theor Phys} {\bf 120}

\bibitem{lvovsky2013novel} Lvovsky Y, Stautner E W, and Zhang T 2013 {\it Supercond. Sci. Technol.} {\bf 26} 093001

\bibitem{bruzzone2018high} Bruzzone P, Fietz W H, Minervini J V, Novikov M, Yanagi N, Zhai Y, and Zheng J 2018 {\it Nucl. Fusion} {\bf 58} 103001

\bibitem{mirhosseini2020superconducting} Mirhosseini M, Sipahigil A, Kalaee M, and Painter O 2020 {\it Nature} {\bf 588} 599

\bibitem{gambetta2017building} Gambetta J M, Chow J M, and Steffen M 2017 {\it NPJ Quantum Inf.} {\bf 3} 2

\bibitem{degen2017quantum} Degen C L, Reinhard F, and Cappellaro P 2017 {\it Rev. Mod. Phys.} {\bf 89} 035002

\bibitem{bardeen1957theory} Bardeen J, Cooper L N, and Schrieffer J R 1957 {\it Phys. Rev.} {\bf 108} 1175

\bibitem {PhysRevLett.130.256002} Ying J J, Liu S Q, Lu Q, Wen X K, Gui Z G, Zhang Y Q, Wang X M, Sun J, and Chen X H 2023 {\it Phys. Rev. Lett.} {\bf 130} 256002

\bibitem {Sun-nature} Sun H L, Huo M W, Hu X W, Li J Y, Liu Z J, Han Y F, Tang L Y, Mao Z Q, Yang P T, Wang B S, Cheng J G, Yao D X, Zhang G M, and Wang M 2023 {\it Nature} {\bf 621} 493

\bibitem {duan2014pressure} Duan D F, Liu Y X, Tian F B, Li D, Huang X L, Zhao Z L, Yu H Y, Liu B B, Tian W J, and Cui T 2014 {\it Sci. Rep.} {\bf 4} 6968

\bibitem {doi:10.1073/pnas.1704505114} Liu H Y, Naumov I I, Hoffmann R, Ashcroft N W, and Hemley R J 2017 {\it Proc. Natl. Acad. Sci. U.S.A.} {\bf 114} 6990

\bibitem {PhysRevLett.119.107001} Peng F, Sun Y, Pickard C J, Needs R J, Wu Q, and Ma Y M 2017 {\it Phys. Rev. Lett.} {\bf 119} 107001

\bibitem {PhysRevB.104.L100504} Gao M, Yan X W, Lu Z Y, and Xiang T 2021 {\it Phys. Rev. B} {\bf 104} L100504

\bibitem {drozdov2019superconductivity} Drozdov A P, Kong P P, Minkov V S, Besedin S P, Kuzovnikov M A, Mozaffari S, Balicas L, Balakirev F F, Graf D E, Prakapenka V B, et al. 2019 {\it Nature} {\bf 569} 528

\bibitem {PhysRevB.104.054501} Shipley A M, Hutcheon M J, Needs R J, and Pickard C J 2021 {\it Phys. Rev. B} {\bf 104} 054501

\bibitem {drozdov2015conventional} Drozdov A P, Eremets M I, Troyan I A, Ksenofontov V, and Shylin S I 2015 {\it Nature} {\bf 525} 73


\bibitem {AImaterials-review} Han X Q, Wang X D, Xu M Y, Feng Z, Yao B W, Guo P J, Gao Z F and Lu Z Y 2025 {\it Chin. Phys. Lett.} {\bf 42} 027403

\bibitem {AI-altermagnetic} Gao Z F, Qu S, Zeng B, Liu Y, Wen J R, Sun H, Guo P J, Lu Z Y 2025 {\it Natl. Sci. Rev.} {\bf nwaf066}

\bibitem {choudhary2022designing} Choudhary K and Garrity K 2022 {\it NPJ Comput. Mater.} {\bf 8} 244

\bibitem {wines2023high} Wines D, Choudhary K, Biacchi A J, Garrity K F and Tavazza F 2023 {\it Nano Lett.} {\bf 23} 969-978

\bibitem {stanev2018machine} Stanev V, Oses C, Kusne A G, Rodriguez E, Paglione J, Curtarolo S and Takeuchi I 2018 {\it NPJ Comput. Mater.} {\bf 4} 29

\bibitem {zhang2022machine} Zhang J, Zhu Z, Xiang X D, Zhang K, Huang S, Zhong C, Qiu H J, Hu K and Lin X 2022 {\it J. Phys. Chem. C} {\bf 126} 8922

\bibitem{menon2022generative} Menon D and Ranganathan R 2022 {\it ACS Omega} {\bf 7} 25958

\bibitem{roter2020predicting}  
Roter B and Dordevic S V 2020 {\it Physica C Supercond} {\bf 575} 1353689

\bibitem{simclp} Han X Q, Xu S S, Feng Z, He R Q, Lu Z Y 2023 {\it Chin. Phys. Lett.} {\bf 40} 027501

\bibitem{wines2023inverse} Wines D, Xie T, and Choudhary K 2023 {\it J. Phys. Chem. Lett.} {\bf 14} 29 6630-6638

\bibitem{xie2022crystal} Xie T, Fu X, Ganea O E, Barzilay R, and Jaakkola T S 2022 {\it ICLR} 25-29 April.

\bibitem{choudhary2020joint} Choudhary K, Garrity K F, Reid A C E, DeCost B, Biacchi A J, Hight Walker A R, Trautt Z, Hattrick-Simpers J, Kusne A G, Centrone A and others 2020 {\it NPJ Comput. Mater.} {\bf 6} 173

\bibitem {alignn2021} Choudhary K and DeCost B 2021 {\it NPJ Comput. Mater.} {\bf 7} 185

\bibitem {supercon50} Cerqueira T F T, Fang Y W, Errea I, Sanna A and Marques M A L 2024 {\it Adv. Funct. Mater.} {\bf 34} 2404043

\bibitem {Sanna2024} Sanna A, Cerqueira T F T, Fang Y W, Errea I, Ludwig A and Marques M A L 2024 {\it NPJ Comput. Mater.} {\bf 10} 44

\bibitem {PhysRevLett.132.166001} Dolui K, Conway L J, Heil C, Strobel T A, Prasankumar R P and Pickard C J 2024 {\it Phys. Rev. Lett.} {\bf 132} 166001

\bibitem {Geng2023} Geng N, Hilleke K P, Zhu L, Wang X, Strobel T A and Zurek E 2023 {\it J. Am. Chem. Soc.} {\bf 145} 1696--1706

\bibitem{ouyang2025strong} Ouyang Z, Yao B W, Guo P J, Gao Z F, Lu Z Y 2025 {\it arXiv} 2501.12222

\bibitem {zhang2023dpa} Zhang D, Liu X, Zhang X, Zhang C, Cai C, Bi H, Du Y, Qin X, Huang J, Li B and others 2024 {\it NPJ Comput. Mater.} {\bf 10} 293

\bibitem {ase-jcp} Yang Y, Jiménez-Negrón O A and Kitchin J R 2021 {\it J. Chem. Phys.} {\bf 154} 234704

\bibitem {jiao2024crystal} Jiao R, Huang W, Lin P, Han J, Chen P, Lu Y and Liu Y 2023 {\it NeurIPS} {\bf 36} 17464--17497

\bibitem {ho2020denoising} Ho J, Jain A and Abbeel P 2020 {\it NeurIPS} {\bf 33} 6840--6851

\bibitem {egnn2021} Satorras V G, Hoogeboom E and Welling M 2021 {\it ICML} {\bf 139} 9323--9332

\bibitem {devlin2018bert} Devlin J, Chang M W, Lee K and Toutanova K 2019 {\it ACL} {\bf 1} 4171

\bibitem {xie2018crystal} Xie T and Grossman J C 2018 {\it Phys. Rev. Lett.} {\bf 120} 145301

\bibitem {ruschendorf1985wasserstein} R{\"u}schendorf L 1985 {\it Probab Theory Relat Fields} {\bf 70} 117--129

\bibitem {chen2019graph} Chen C, Ye W, Zuo Y, Zheng C, Ong S P 2019 {\it Chem. Mater} {\bf 31} 3564--3572

\bibitem {Zeni2025} Zeni C, Pinsler R, Zügner D, Fowler A, Horton M, Fu X, Wang Z, Shysheya A, Crabbé J, Ueda S, Sordillo R, Sun L, Smith J, Nguyen B, Schulz H, Lewis S, Huang C, Lu Z, Zhou Y, Yang H, Hao H, Li J, Yang C, Li W, Tomioka R, Xie T 2025 {\it Nature}

\bibitem{song2021Score-based} Song Y, Sohl-Dickstein J, Kingma DP, Kumar A, Ermon S, Poole B 2021 {\it ICLR}

\bibitem {jain2020materials} Jain A, Montoya J, Dwaraknath et al. 2020 {\it Handbook of Materials Modeling} {\bf 1751--1784}

\bibitem {liu1989limited} Liu D C, Nocedal J 1989 {\it Math Program} {\bf 45} 503--528

\bibitem {SchNet2017} Schütt K, Kindermans P J, Sauceda F H, Chmiela S, Tkatchenko A, Müller K R 2017 {\it NeurIPS} {\bf 30} 991--1001

\bibitem {merchant2023scaling} Merchant A, Batzner S, Schoenholz S S, Aykol M, Cheon G, Cubuk E D 2023 {\it Nature} {\bf 624} 80--85

\bibitem {PhysRevB.6.2577} Allen P B 1972 {\it Phys. Rev. B} {\bf 6} 2577--2579

\bibitem {PhysRevB.12.905} Allen P B, Dynes R C 1975 {\it Phys. Rev. B} {\bf 12} 905--922

\bibitem {concdvae} Ye C Y, Weng H M, Wu Q S 2024 {\it Comput. Mater. Today} {\bf 1} 100003

\bibitem {Antunes2024} Antunes L M, Butler K T, Grau-Crespo R 2024 {\it Nat. Commun.} {\bf 15} 10570


\end{thebibliography}


\end{document}